\documentclass[12pt]{article}

\usepackage[utf8]{inputenc}
\usepackage{charter}
\usepackage{fullpage}
\usepackage{hyperref}
\usepackage{graphicx}
\usepackage{lipsum}
\usepackage[sort&compress,numbers]{natbib}
\usepackage{hyperref}
\hypersetup{hidelinks}
\usepackage[total={6.5in,9in},top=1in,headsep=0.1in,headheight=1in]{geometry}

\usepackage{booktabs,amsmath,amssymb,multirow}

\newcommand{\HVPlatticeref}{Chakraborty:2017tqp,Borsanyi:2017zdw,Blum:2018mom,Giusti:2019xct,Shintani:2019wai,FermilabLattice:2019ugu,Gerardin:2019rua,Aubin:2019usy,Giusti:2019hkz}

\newcommand{\HVPtotalref}{Davier:2017zfy,Keshavarzi:2018mgv,Colangelo:2018mtw,Hoferichter:2019gzf,Davier:2019can,Keshavarzi:2019abf,Kurz:2014wya}

\newcommand{\HLbLlatticeref}{Blum:2019ugy}

\newcommand{\HLbLlatticemethods}{Blum:2014oka,Green:2015mva,Blum:2015gfa,Blum:2016lnc,Asmussen:2016lse,Blum:2017cer,Asmussen:2019act}

\newcommand{\HLbLref}{Melnikov:2003xd,Masjuan:2017tvw,Colangelo:2017fiz,Hoferichter:2018kwz,Gerardin:2019vio,Bijnens:2019ghy,Colangelo:2019uex,Pauk:2014rta,Danilkin:2016hnh,Jegerlehner:2017gek,Knecht:2018sci,Eichmann:2019bqf,Roig:2019reh}

\newcommand{\QEDref}{Aoyama:2012wk,Aoyama:2019ryr}

\newcommand{\EWref}{Czarnecki:2002nt,Gnendiger:2013pva}

\newcommand{\HLbLcombref}{Melnikov:2003xd,Masjuan:2017tvw,Colangelo:2017fiz,Hoferichter:2018kwz,Gerardin:2019vio,Bijnens:2019ghy,Colangelo:2019uex,Pauk:2014rta,Danilkin:2016hnh,Jegerlehner:2017gek,Knecht:2018sci,Eichmann:2019bqf,Roig:2019reh,Blum:2019ugy}

\newcommand{\HLbLtotalref}{Melnikov:2003xd,Masjuan:2017tvw,Colangelo:2017fiz,Hoferichter:2018kwz,Gerardin:2019vio,Bijnens:2019ghy,Colangelo:2019uex,Pauk:2014rta,Danilkin:2016hnh,Jegerlehner:2017gek,Knecht:2018sci,Eichmann:2019bqf,Roig:2019reh,Blum:2019ugy,Colangelo:2014qya}

\newcommand{\SMref}{Aoyama:2012wk,Aoyama:2019ryr,Czarnecki:2002nt,Gnendiger:2013pva,Davier:2017zfy,Keshavarzi:2018mgv,Colangelo:2018mtw,Hoferichter:2019gzf,Davier:2019can,Keshavarzi:2019abf,Kurz:2014wya,Melnikov:2003xd,Masjuan:2017tvw,Colangelo:2017fiz,Hoferichter:2018kwz,Gerardin:2019vio,Bijnens:2019ghy,Colangelo:2019uex,Blum:2019ugy,Colangelo:2014qya}

\newcommand{\HVPexpref}{Bai:1999pk,Akhmetshin:2000ca,Akhmetshin:2000wv,Achasov:2000am,Bai:2001ct,Achasov:2002ud,Akhmetshin:2003zn,Aubert:2004kj,Aubert:2005eg,Aubert:2005cb,Aubert:2006jq,Aulchenko:2006na,Achasov:2006vp,Akhmetshin:2006wh,Akhmetshin:2006bx,Akhmetshin:2006sc,Aubert:2007ur,Aubert:2007ef,Aubert:2007uf,Aubert:2007ym,Akhmetshin:2008gz,Ambrosino:2008aa,Ablikim:2009ad,Aubert:2009ad,Ambrosino:2010bv,Lees:2011zi,Lees:2012cr,Lees:2012cj,Babusci:2012rp,Akhmetshin:2013xc,Lees:2013ebn,Lees:2013uta,Lees:2014xsh,Achasov:2014ncd,Aulchenko:2014vkn,Akhmetshin:2015ifg,Ablikim:2015orh,Shemyakin:2015cba,Anashin:2015woa,Achasov:2016bfr,Achasov:2016lbc,TheBaBar:2017aph,CMD-3:2017tgb,TheBaBar:2017vzo,Kozyrev:2017agm,Anastasi:2017eio,Achasov:2017vaq,Xiao:2017dqv,TheBaBar:2018vvb,Anashin:2018vdo,Achasov:2018ujw,Lees:2018dnv,CMD-3:2019ufp}

\newcommand{\HLbLexpref}{Behrend:1990sr,Gronberg:1997fj,Acciarri:1997yx,Achard:2001uu,Achard:2007hm,Arnaldi:2009aa,Aubert:2009mc,BABAR:2011ad,Berghauser:2011zz,Uehara:2012ag,Babusci:2012ik,Aguar-Bartolome:2013vpw,Ablikim:2015wnx,Masuda:2015yoh,Arnaldi:2016pzu,Adlarson:2016hpp,Adlarson:2016ykr,TheNA62:2016fhr,BaBar:2018zpn,PrimEx-II:2020jwd}

\newcommand{\amuSM}{\ensuremath{a_\mu^\text{SM}}}

\newcommand{\amuexp}{\ensuremath{a_\mu^\text{exp}}}

\newcommand{\amuHVPLO}{\ensuremath{a_\mu^\text{HVP, LO}}}

\newcommand{\GeV}{\,\text{GeV}}

\newcommand{\Order}{\mathcal{O}}

\newcommand\snowmass{
\begin{center}
  \rule[-0.2in]{\hsize}{0.01in}\\
  \rule{\hsize}{0.01in}\\
  \vskip 0.1in
  Submitted to the Proceedings of the US Community Study\\ 
  on the Future of Particle Physics (Snowmass 2021)\\
  \rule{\hsize}{0.01in}\\
  \rule[+0.2in]{\hsize}{0.01in}\\[-2em]
\end{center}
}

\usepackage[firstpage=true]{background}
\backgroundsetup{contents={\parbox{6.5in}{\snowmass}}, scale=1,placement=top,opacity=1,color=black,position={3.25in,1.2in}}

\usepackage{fancyhdr}
\fancypagestyle{plain}{%
  \fancyhf{}%
  \fancyhead[C]{}
  \fancyfoot[C]{\thepage}
}

\fancypagestyle{empty}{%
  \fancyhf{}%
  \fancyhead[C]{{\it Prospects for precise predictions of $a_\mu$ in the SM}}
  \fancyfoot[C]{\thepage}
}
\title{\vspace{-1.5cm}{\noindent{\footnotesize{FERMILAB-CONF-22-236-T}}}\hfill{\footnotesize{LTH 1303}}\hfill{\footnotesize{MITP-22-030}}\\[0.7cm]
Prospects for precise predictions of $a_\mu$ in the Standard Model}
\date{}

\usepackage{authblk}

\author[1]{G.~Colangelo}
\author[2]{M.~Davier}
\author[3,4]{A.~X.~El-Khadra}
\author[1]{M.~Hoferichter}
\author[5]{C.~Lehner}
\author[6]{L.~Lellouch}
\author[7]{T.~Mibe}
\author[8]{B.~L.~Roberts}
\author[9]{T.~Teubner}
\author[10,11]{H.~Wittig}
\author[12]{B.~Ananthanarayan}
\author[13]{A.~Bashir}
\author[14]{J.~Bijnens}
\author[15,16]{T.~Blum}
\author[17]{P.~Boyle}
\author[18]{N.~Bray-Ali}
\author[19]{I.~Caprini}
\author[20]{C.~M.~Carloni Calame}
\author[21]{O.~Cat\`a}
\author[1]{M.~C\`e}
\author[6]{J.~Charles}
\author[22]{N.~H.~Christ}
\author[23]{F.~Curciarello}
\author[10]{I.~Danilkin}
\author[24]{D.~Das}
\author[10]{O.~Deineka}
\author[25]{M.~Della Morte}
\author[10]{A.~Denig}
\author[26]{C.~E.~DeTar}
\author[27]{C.~A.~Dominguez}
\author[28]{G.~Eichmann}
\author[29]{C.~S.~Fischer}
\author[6]{A.~G\'erardin}
\author[5]{D.~Giusti}
\author[30,31,32]{M.~Golterman}
\author[33]{Steven~Gottlieb}
\author[34]{V.~G\"ulpers}
\author[10,35]{F.~Hagelstein}
\author[36,37]{M.~Hayakawa}
\author[1]{N.~Hermansson-Truedsson}
\author[1]{B.-L.~Hoid}
\author[38]{S.~Holz}
\author[17,16]{T.~Izubuchi}
\author[39,40]{A.~J\"uttner}
\author[41]{A.~Keshavarzi}
\author[6]{M.~Knecht}
\author[4]{A.~S.~Kronfeld}
\author[38]{B.~Kubis}
\author[42]{A.~Kup\'s\'c}
\author[3]{S.~Lahert}
\author[43]{K.~F.~Liu}
\author[44]{J.~L\"udtke}
\author[3]{M.~Lynch}
\author[45]{B.~Malaescu}
\author[46,47]{K.~Maltman}
\author[17]{W.~Marciano}
\author[48]{M.~K.~Marinkovi\'c}
\author[31,32]{P.~Masjuan}
\author[10,11]{H.~B.~Meyer}
\author[49]{S.~E.~M\"uller}
\author[50]{E.~T.~Neil}
\author[51]{M.~Passera}
\author[52]{M.~Pepe}
\author[31,32]{S.~Peris}
\author[53]{A.~A.~Petrov}
\author[44]{M.~Procura}
\author[54]{K.~Raya}
\author[55]{A.~Rebhan}
\author[56]{A.~Risch}
\author[2]{A.~Rodr\'iguez-S\'anchez}
\author[57]{P.~Roig}
\author[31]{P.~S\'anchez-Puertas}
\author[58]{S.~Simula}
\author[59,35]{P.~Stoffer}
\author[60]{F.~M.~Stokes}
\author[61]{R.~Sugar}
\author[25]{J.~T.~Tsang}
\author[4]{R.~S.~van de Water}
\author[26]{A.~Vaquero  Avil\'es-Casco}
\author[62]{G.~Venanzoni}
\author[10]{G.~M.~von Hippel}
\author[2]{Z.~Zhang}

\affil[1]{Albert Einstein Center for Fundamental Physics, Institute for Theoretical Physics, University of Bern, Sidlerstrasse 5, 3012 Bern, Switzerland}
\affil[2]{IJCLab, Universit\'e Paris-Saclay and CNRS/IN2P3, 91405 Orsay, France}
\affil[3]{Department of Physics and Illinois Center for Advanced Studies of the Universe, University of Illinois, Urbana, IL 61801, USA}
\affil[4]{Particle Physics Department, Theory Division, Fermi National Accelerator Laboratory, Batavia, IL 60510, USA}
\affil[5]{Universit\"at Regensburg, Fakult\"at f\"ur Physik, Universit\"atsstra\ss e 31, 93040 Regensburg, Germany}
\affil[6]{Aix Marseille Univ, Universit\'{e} de Toulon, CNRS, CPT, Marseille, France}
\affil[7]{Institute of Particle and Nuclear Studies, High Energy Accelerator Research Organization (KEK), Tsukuba 305-0801, Japan}
\affil[8]{Department of Physics, Boston University, Boston, MA 02215, USA}
\affil[9]{Department of Mathematical Sciences, University of Liverpool, Liverpool L69 3BX, United Kingdom}
\affil[10]{PRISMA$^+$ Cluster of Excellence and Institute for Nuclear Physics,
 Johannes Gutenberg University of Mainz, 55099 Mainz, Germany}
 \affil[11]{Helmholtz Institute Mainz, 55099 Mainz, Germany and GSI Helmholtzzentrum f\"ur Schwerionenforschung, 64291 Darmstadt, Germany}
 \affil[12]{Centre for High Energy Physics, Indian Institute of Science,	Bangalore 560 012, India}
 \affil[13]{Instituto de F\'isica y Matem\'aticas, Universidad Michoacana de San Nicol\'as de Hidalgo, Morelia, Michoac\'an 58040, M\'exico}
 \affil[14]{Department of Astronomy and Theoretical Physics, Lund University, S\"olvegatan 14A, 22362 Lund, Sweden}
 \affil[15]{Department of Physics, 196 Auditorium Road, Unit 3046, University of Connecticut, Storrs, CT 06269-3046, USA}
 \affil[16]{RIKEN BNL Research Center,  Brookhaven National Laboratory, Upton, NY 11973, USA}
  \affil[17]{Physics Department, Brookhaven National Laboratory, Upton, NY 11973, USA}
 \affil[18]{Department of Physical Sciences, Mount Saint Mary's University-Los Angeles, USA}
 \affil[19]{Horia Hulubei National Institute for Physics and Nuclear Engineering, P.O.B.\ MG-6, 077125 Bucharest-Magurele, Romania}
 \affil[20]{Istituto Nazionale di Fisica Nucleare (INFN), Sezione di Pavia,
 Via A.\ Bassi 6, 27100 Pavia, Italy}
 \affil[21]{Center for Particle Physics Siegen (CPPS), Theoretische Physik 1, Universit\"at Siegen, Walter-Flex-Str.~3, 57068 Siegen, Germany}
 \affil[22]{Department of Physics, Columbia University,New York, NY 10027, USA}
 \affil[23]{Department of Physics, University of Calabria, Via P.~Bucci, Arcavacata di Rende (CS), Italy}
 \affil[24]{Center for Computational Natural Sciences and Bioinformatics, International Institute of Information Technology, Hyderabad,	Prof.~C R Rao Road, Gachibowli, Hyderabad 500032, Telangana, India}
  \affil[25]{CP3-Origins and IMADA, University of Southern Denmark, Campusvej 55, 5230 Odense M, Denmark}
  \affil[26]{Department of Physics and Astronomy, University of Utah, Salt Lake City, UT 84112, USA}
 \affil[27]{Centre for Theoretical and Mathematical Physics, and Department of Physics, University of Cape
Town, Rondebosch 7700, South Africa}
 \affil[28]{LIP Lisboa, Av.\ Prof.\ Gama Pinto 2, 1649-003 Lisboa, Portugal}
 \affil[29]{Institute for Theoretical Physics, Justus-Liebig University, Heinrich-Buff-Ring 16, 35392 Gie\ss en, Germany}
 \affil[30]{Department of Physics and Astronomy, San Francisco State University, San Francisco, CA 94132, USA}
 \affil[31]{Grup de F\'{\i}sica Te\`orica, Departament de F\'{\i}sica, Universitat Aut\`onoma de Barcelona, 08193~Bellaterra (Barcelona), Spain}
 \affil[32]{Institut de F{\'i}sica d'Altes Energies (IFAE) and The Barcelona Institute of Science and Technology, Universitat Aut{\'o}noma de Barcelona, 08193~Bellaterra (Barcelona), Spain}
 \affil[33]{Department of Physics, Indiana University, Bloomington, IN 47405, USA}
 \affil[34]{School of Physics and Astronomy, University of Edinburgh, Edinburgh EH9 3FD, United Kingdom}
  \affil[35]{Paul Scherrer Institut, 5232 Villigen PSI, Switzerland}
 \affil[36]{Department of Physics, Nagoya University, Nagoya 464-8602, Japan}
 \affil[37]{Nishina Center, RIKEN, Wako 351-0198, Japan}
 \affil[38]{Helmholtz-Institut f\"ur Strahlen- und Kernphysik (Theorie) and Bethe Center for Theoretical Physics, Universit\"at Bonn, 53115 Bonn, Germany}
 \affil[39]{Theoretical Physics Department, CERN, 1211 Geneva 23, Switzerland}
 \affil[40]{School of Physics and Astronomy, University of Southampton, Southampton, SO17 1BJ, UK}
 \affil[41]{Department of Physics and Astronomy, The University of Manchester, Manchester M13 9PL, United Kingdom}
 \affil[42]{Uppsala University and National Centre for Nuclear Research, NCBJ, Box 516, 75120 Uppsala, Sweden}
 \affil[43]{Department of Physics and Astronomy, University of Kentucky,  Lexington, KY 40506, USA}
  \affil[44]{University of Vienna, Faculty of Physics, Boltzmanngasse 5, 1090 Wien, Austria}
 \affil[45]{LPNHE, Sorbonne Universit\'e, Universit\'e Paris Cit\'e, CNRS/IN2P3, Paris, France}
 \affil[46]{Mathematics and Statistics, York University, Toronto, ON, Canada}
 \affil[47]{CSSM, University of Adelaide, Adelaide, SA, Australia}
 \affil[48]{Institute for Theoretical Physics, ETH Zurich, Wolfgang-Pauli-Str.~27, 8093 Zurich, Switzerland}
 \affil[49]{Helmholtz-Zentrum Dresden-Rossendorf, Bautzner Landstra\ss e 400, 01328 Dresden, Germany}
 \affil[50]{Department of Physics, University of Colorado, Boulder, CO 80309, USA}
 \affil[51]{Istituto Nazionale di Fisica Nucleare (INFN), Sezione di Padova, Via Francesco Marzolo 8, 35131 Padova, Italy}
 \affil[52]{Istituto Nazionale di Fisica Nucleare (INFN), Sezione di Milano-Bicocca	Piazza della Scienza 3, I-20126, Milano, Italy}
 \affil[53]{Department of Physics and Astronomy, Wayne State University, Detroit, MI 48201, USA}
 \affil[54]{Departamento de F\'isica Te\'orica y del Cosmos, Universidad de Granada	18071, Granada, Spain}
 \affil[55]{Institute for Theoretical Physics, Technische Universit\"at Wien, Wiedner Hauptstr.~8-10, 1040 Vienna, Austria}
 \affil[56]{John von Neumann-Institut f\"ur Computing NIC, Deutsches Elektronen-Synchrotron DESY, Platanenallee 6, 15738 Zeuthen, Germany}
 \affil[57]{Departamento de F\'isica, Centro de Investigaci\'on y de Estudios Avanzados del Instituto Polit\'ecnico Nacional, Apdo.\ Postal 14-740, 07000~Ciudad de M\'exico D.~F., M\'exico}
 \affil[58]{Istituto Nazionale di Fisica Nucleare (INFN), Sezione di Roma Tre, Via della Vasca Navale 84, 00146 Roma, Italy}
 \affil[59]{Physik-Institut, Universit\"at Z\"urich,	Winterthurerstrasse 190, 8057 Z\"urich, Switzerland}
 \affil[60]{J\"ulich Supercomputing Centre, Forschungszentrum J\"ulich, 52428 J\"ulich, Germany}
 \affil[61]{Physics Department, University of California Santa Barbara, Santa Barbara, CA 93106}
 \affil[62]{Istituto Nazionale di Fisica Nucleare (INFN), Sezione di Pisa, Largo Bruno Pontecorvo 3, 56127 Pisa, Italy}

\begin{document}

\maketitle

\begin{abstract}

We discuss the prospects for improving the precision on the hadronic corrections to the anomalous magnetic moment of the muon, and the plans of the Muon $g-2$ Theory Initiative to update the Standard Model prediction. 
\end{abstract}

\section{Introduction}

The Run-1 result of the Fermilab $g-2$ experiment~\cite{Muong-2:2021ojo,Muong-2:2021ovs,Muong-2:2021xzz,Muong-2:2021vma} confirmed the earlier BNL measurement~\cite{Muong-2:2006rrc}, resulting in a $4.2\sigma$ tension of the combined experimental value with respect to the recommendation by the White Paper~\cite{Aoyama:2020ynm} of the Muon $g-2$ Theory Initiative~\cite{MuonInitiative} (reflecting discussions at a series of workshops~\cite{FNAL2017,KEK2018,UConn2018,Mainz2018,INT2019}). 
In Table~\ref{tab:summary} we reproduce the status of the Standard-Model (SM) prediction as presented therein, as reference point for the prospects of future improvements. With results from subsequent runs of the Fermilab experiment expected soon, poised to reduce the experimental uncertainty by more than another factor of $2$~\cite{Muong-2:2015xgu}, as well as future $g-2$ experiments at J-PARC~\cite{Abe:2019thb} and, potentially, PSI~\cite{Aiba:2021bxe} and Fermilab~\cite{FutMuon2021}, it is clear that theory needs to be improved concurrently.  

\begin{table}[t]
\begin{centering}
\small
	\begin{tabular}{l  r l }
	\toprule
	   Contribution  & Value   $\times 10^{11}$ & References\\ \midrule
	   Experiment (E821 + E989)                     &$ 116\, 592\, 061(41)$ & Refs.~\cite{Muong-2:2006rrc,Muong-2:2021ojo}  \\\midrule
	HVP LO ($e^+e^-$)  & $ 6931(40) $  &  Refs.~\cite{Davier:2017zfy,Keshavarzi:2018mgv,Colangelo:2018mtw,Hoferichter:2019gzf,Davier:2019can,Keshavarzi:2019abf}\\
        HVP NLO ($e^+e^-$) & $-98.3(7) $ & Ref.~\cite{Keshavarzi:2019abf}\\
        HVP NNLO ($e^+e^-$)  & $ 12.4(1) $ & Ref.~\cite{Kurz:2014wya}\\
        HVP LO (lattice, $udsc$)   & $7116 (184)$ & Refs.~\cite{\HVPlatticeref}\\
HLbL (phenomenology)  & $92(19)$ & Refs.~\cite{\HLbLref}\\
HLbL NLO (phenomenology)  & $2(1)$ & Ref.~\cite{Colangelo:2014qya}\\
HLbL (lattice, $uds$)  & $79(35)$ & Ref.~\cite{\HLbLlatticeref}\\
HLbL (phenomenology + lattice) & $90(17)$ & Refs.~\cite{\HLbLcombref}\\\midrule
QED                &  $116\,584\,718.931(104)$  & Refs.~\cite{\QEDref}\\
Electroweak      & $153.6(1.0)$   & Refs.~\cite{\EWref}\\
HVP ($e^+e^-$, LO + NLO + NNLO)  & $6845(40)$ & Refs.~\cite{\HVPtotalref} \\
HLbL (phenomenology + lattice + NLO)  & $92(18)$ & Refs.~\cite{\HLbLtotalref}\\ 
        Total SM Value   & $ 116\, 591\, 810(43) $  & Refs.~\cite{\SMref}\\
        Difference:    $\Delta a_\mu:=\amuexp - \amuSM$  & $ 251(59) $ & \\
        \bottomrule
	\end{tabular}
	\caption{Summary of the contributions to $\amuSM$, as compiled in Ref.~\cite{Aoyama:2020ynm}, except for the update of the experimental number to the average of E821 and the first Run of E989. The first block gives the main results for the hadronic contributions as well as the combined result for HLbL scattering from phenomenology and lattice QCD available at the time of Ref.~\cite{Aoyama:2020ynm}. 
	The second block summarizes the quantities entering the final recommendation for the SM contribution, in particular, the total HVP contribution, evaluated from $e^+e^-$ data, and the total HLbL number.     
	The HVP evaluation is mainly based on the experimental Refs.~\cite{\HVPexpref}.
	In addition, the HLbL evaluation uses experimental input from Refs.~\cite{\HLbLexpref}.  The lattice QCD calculation of the HLbL contribution builds on crucial methodological advances from Refs.~\cite{\HLbLlatticemethods}. Finally, the QED value uses the fine-structure constant obtained from atom-interferometry measurements of the Cs atom~\cite{Parker:2018vye}, and is affected by the tension with the more recent Rb result~\cite{Morel:2020dww} only at a level irrelevant for $\amuSM$. Mixed leptonic and hadronic corrections enter at the same order $\Order(\alpha^4)$ as HVP NNLO and HLbL NLO, but have been estimated as $\lesssim 1\times 10^{-11}$~\cite{Hoferichter:2021wyj}. }
\label{tab:summary}
\end{centering}
\end{table}

This is particularly pressing given the tension between hadronic vacuum polarization (HVP) extracted from $e^+e^-\to\text{hadrons}$ cross section data, upon which the final value from Ref.~\cite{Aoyama:2020ynm} is based, and the recent lattice calculation by the BMW collaboration~\cite{Borsanyi:2020mff}. Here the most urgent task is to scrutinize the result of Ref.~\cite{Borsanyi:2020mff} in detailed comparisons with lattice results of commensurate precision obtained in independent calculations by other lattice collaborations. As discussed in Sec.~\ref{sec:latticeHVP}, such calculations are forthcoming. If the tensions persist, their phenomenological consequences must also be explored~\cite{Crivellin:2020zul,Keshavarzi:2020bfy,Malaescu:2020zuc,Colangelo:2020lcg,Ce:2022eix} (see Sec.~\ref{sec:outlook}).    
Moreover, also the hadronic light-by-light (HLbL) contribution needs to be further improved to meet the final precision $\Delta a_\mu^\text{E989}=16\times 10^{-11}$ projected for the Fermilab experiment~\cite{Muong-2:2015xgu}. 

\begin{figure}[htb]
    \centering
    \includegraphics[width=0.5\textwidth]{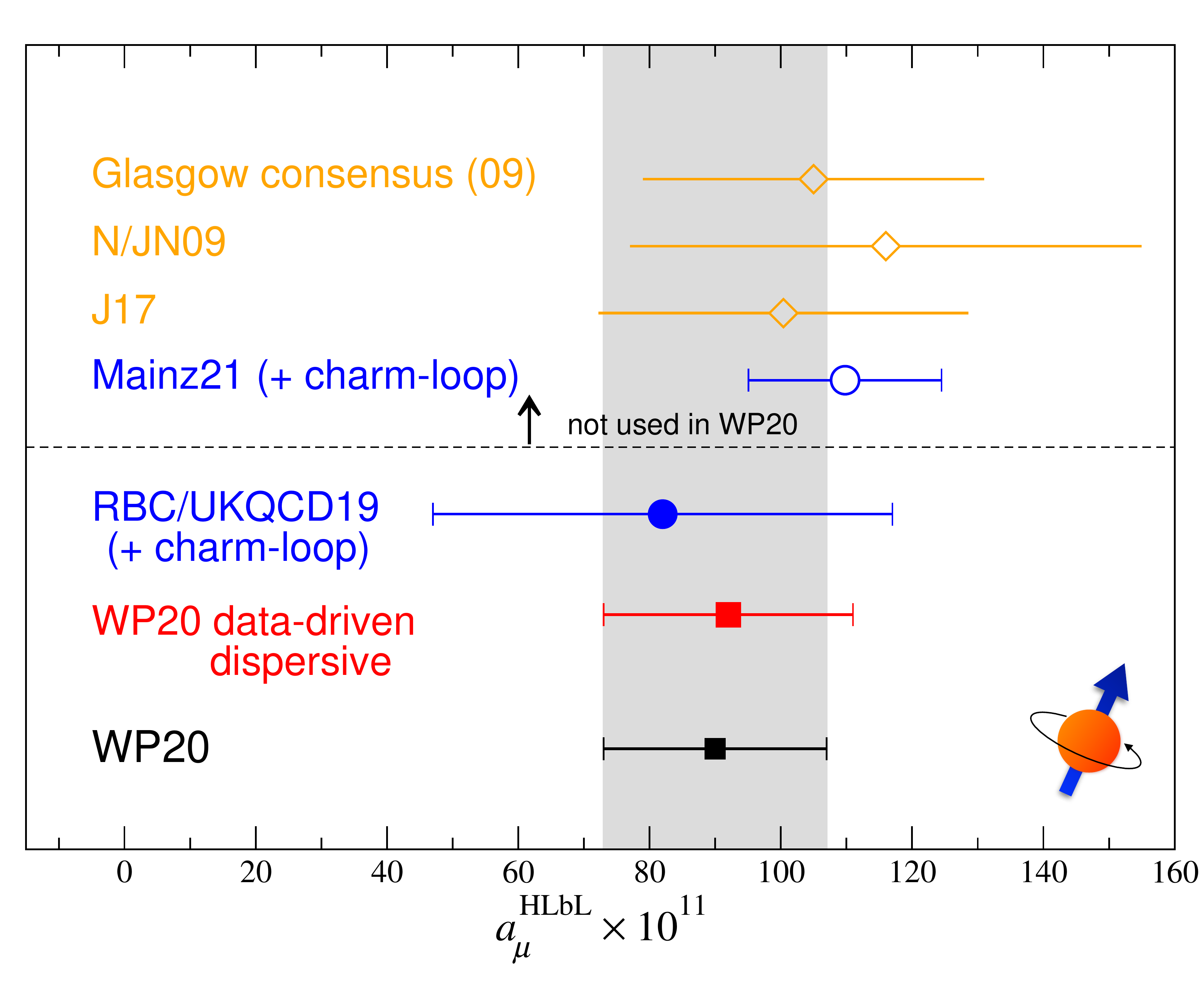}\hfill
    \includegraphics[width=0.5\textwidth]{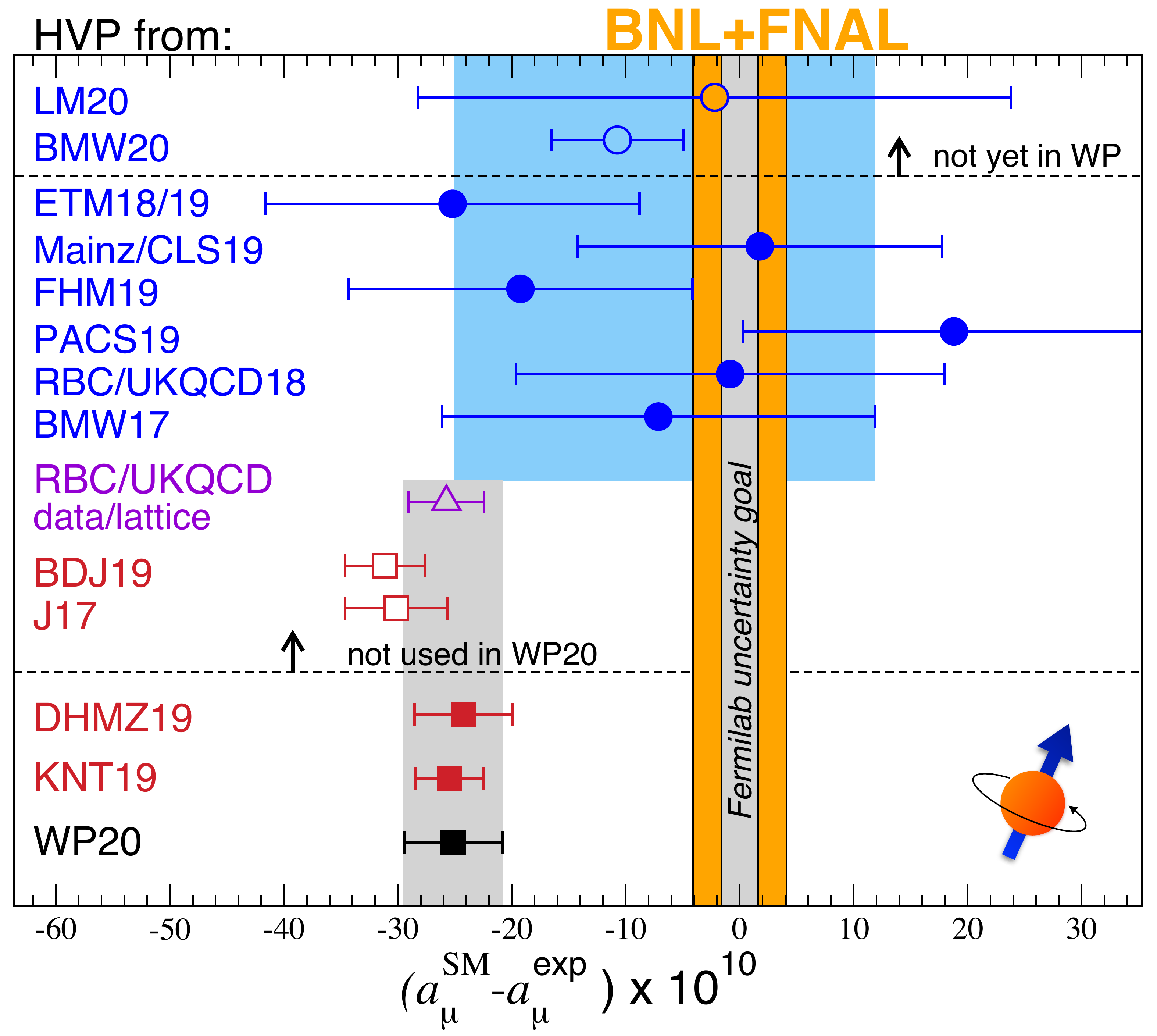}
    \caption{Left: Comparison of HLbL evaluations, as quoted in Ref.~\cite{Aoyama:2020ynm}, to earlier estimates~\cite{Prades:2009tw,Nyffeler:2009tw, Jegerlehner:2009ry,Jegerlehner:2017gek} (orange) and a more recent lattice calculation~\cite{Chao:2021tvp} (open blue).
    Right: Comparison of theoretical predictions of $a_\mu$ with experiment~\cite{Muong-2:2021ojo,Muong-2:2006rrc} (orange band), adapted from Ref.~\cite{Aoyama:2020ynm}. Each data point represents a different evaluation of leading-order HVP, to which the remaining SM contributions, as given in Ref.~\cite{Aoyama:2020ynm}, have been added. Red squares show data-driven results~\cite{Jegerlehner:2017gek,Benayoun:2019zwh,Davier:2019can,Keshavarzi:2019abf}; filled blue circles indicate lattice-QCD calculations that were taken into account in the WP20 lattice average~\cite{Borsanyi:2017zdw,Blum:2018mom,Giusti:2019xct,Shintani:2019wai,FermilabLattice:2019ugu,Gerardin:2019rua,Giusti:2019hkz}, while the open ones show results published after the deadline for inclusion in that average~\cite{Borsanyi:2020mff,Lehner:2020crt}; the purple triangle gives a hybrid of the two~\cite{Blum:2018mom}.
        The SM prediction of Ref.~\cite{Aoyama:2020ynm} is shown as the black square and gray band.}
    \label{fig:compare}
\end{figure}

 A comparison of published results for HVP and HLbL, including those that were published after the March 2020 deadline, is shown in Fig.~\ref{fig:compare}.  In this contribution, we briefly review the current status from data-driven evaluations and from lattice QCD for both quantities 
and discuss future prospects as well as future plans of the Muon $g-2$ Theory Initiative.       

\section{Data-driven evaluations of HVP}
\label{sec:dataHVP}

\begin{table}[t]
\small
	\centering
	\begin{tabular}{c r r r}
	\toprule
	  & Ref.~\cite{Davier:2019can} & Ref.~\cite{Keshavarzi:2019abf} & Difference\\\midrule
	 $\pi^+\pi^-$ & $507.85(0.83)(3.23)(0.55)$ & $504.23(1.90)$ & $3.62$\\
	 $\pi^+\pi^-\pi^0$ & $46.21(0.40)(1.10)(0.86)$ & $46.63(94)$ & $-0.42$\\
	 $\pi^+\pi^-\pi^+\pi^-$ & $13.68(0.03)(0.27)(0.14)$ & $13.99(19)$ & $-0.31$\\
	 $\pi^+\pi^-\pi^0\pi^0$ & $18.03(0.06)(0.48)(0.26)$ & $18.15(74)$ & $-0.12$\\
	 $K^+K^-$ & $23.08(0.20)(0.33)(0.21)$ & $23.00(22)$ & $0.08$\\
	 $K_SK_L$ & $12.82(0.06)(0.18)(0.15)$ & $13.04(19)$ & $-0.22$\\
	 $\pi^0\gamma$ & $4.41(0.06)(0.04)(0.07)$ & $4.58(10)$ & $-0.17$\\\midrule
	 Sum of the above  & $626.08(0.95)(3.48)(1.47)$ & $623.62(2.27)$ & $2.46$\\\midrule
	 $[1.8,3.7]\GeV$ (without $c \bar c$) & $33.45(71)$ & $34.45(56)$ & $-1.00$\\
	 $J/\psi$, $\psi(2S)$ & $7.76(12)$ & $7.84(19)$ & $-0.08$ \\
	 $[3.7,\infty)\GeV$ & $17.15(31)$ & $16.95(19)$ & $0.20$\\\midrule
	 Total $\amuHVPLO$ & $694.0(1.0)(3.5)(1.6)(0.1)_\psi(0.7)_\textrm{DV+QCD}$ & $692.8(2.4)$ & $1.2$\\ 
	\bottomrule
	\end{tabular}
	\caption{Comparison of selected exclusive-mode contributions to $\amuHVPLO$ from Refs.~\cite{Davier:2019can,Keshavarzi:2019abf}, for the energy range $\leq 1.8\GeV$, in units of $10^{-10}$, see Ref.~\cite{Aoyama:2020ynm} for details.}
\label{tab:KNT_DHMZ}
\end{table}

The data-driven evaluation of HVP relies on the master formula from Refs.~\cite{Bouchiat:1961lbg,Brodsky:1967sr}, a dispersion relation that relates the leading-order HVP contribution $\amuHVPLO$ to the total cross section for $e^+e^-\to\text{hadrons}$.\footnote{The cross section is defined photon-inclusively, see Ref.~\cite{Aoyama:2020ynm}, i.e., while $\amuHVPLO$ is $\Order(\alpha^2)$, it contains, by definition, one-photon-irreducible
contributions of order $\Order(\alpha^3)$. This convention matches the one used in lattice-QCD calculations.} The main challenges in converting the available data~\cite{\HVPexpref} to the corresponding HVP integral include the combination of data sets in the presence of tensions in the data base and the propagation and assessment of the resulting uncertainties. For illustration, the contributions of the main exclusive channels and the inclusive region from the compilations of Refs.~\cite{Davier:2019can,Keshavarzi:2019abf} are shown in Table~\ref{tab:KNT_DHMZ}. 

In Ref.~\cite{Aoyama:2020ynm} a conservative merging procedure was defined 
to obtain a realistic assessment of these underlying uncertainties. The procedure accounts for tensions among the data sets, for differences in methodologies in the combination of experimental inputs, for correlations between systematic errors, and includes constraints from unitarity and analyticity~\cite{Colangelo:2018mtw,Ananthanarayan:2018nyx,Davier:2019can,Hoferichter:2019gzf}. Further, the next-to-leading-order calculation from Ref.~\cite{Campanario:2019mjh} suggests that radiative corrections are under control at this level. 

Recent developments in the data-driven HVP evaluation that are not yet reflected in 
the recommendation from Ref.~\cite{Aoyama:2020ynm} include the crucial $2\pi$ channel (new data from SND~\cite{SND:2020nwa} and covariance matrix from BESIII~\cite{Ablikim:2015orh}) as well as new data for $e^+e^-\to 3\pi$~\cite{BESIII:2019gjz,BABAR:2021cde}, the second-largest channel both in absolute value and error, see Table~\ref{tab:KNT_DHMZ}.  Moreover, unitarity and analyticity constraints have been analyzed for the $\pi^0\gamma$~\cite{Hoid:2020xjs} and $\bar K K$ channels~\cite{Stamen:2022uqh}. However, as of now, none of these developments indicate significant changes compared to the situation described in Ref.~\cite{Aoyama:2020ynm}. 

In going forward, new data in the critical $2\pi$ channel at the same level of precision as BaBar~\cite{Aubert:2009ad,Lees:2012cj} and KLOE~\cite{Ambrosino:2008aa,Ambrosino:2010bv,Babusci:2012rp,Anastasi:2017eio} are required. 
Such data are expected in the coming years from BaBar, CMD-3, BESIII, and Belle II, besides new data for other channels as well. To credibly resolve the existing tensions, especially for the $2\pi$ channel, blind analyses are paramount. Finally, for the success of this program the development of Monte Carlo generators at NNLO accuracy is necessary, see Refs.~\cite{Abbiendi:2022liz,WorkingGrouponRadiativeCorrections:2010bjp}. 

The precision that can be obtained for data-driven evaluations of HVP strongly depends on whether or not the present tension between the BABAR and KLOE experiments, see Ref.~\cite{Aoyama:2020ynm}, can be resolved with the upcoming advent of new $2\pi$ analyses. If the answer to that question is affirmative, a precision of $0.3\%$ seems feasible by 2025.

\section{Lattice QCD calculations of HVP}
\label{sec:latticeHVP}

HVP can also be computed from first principles in QCD using a non-perturbative lattice regulator. Calculations are performed in Euclidean space, and the HVP contribution is computed by a weighted integration of the correlation functions over Euclidean time. In lattice QCD calculations, the total HVP is obtained from a sum over all quark-flavors and includes connected and disconnected contractions. Almost all gauge-field ensembles generated by the various lattice collaborations to date include light sea quarks in the isospin symmetric limit ($m_u=m_d$). Hence, strong isospin-breaking corrections must be computed alongside the $\Order(\alpha^3)$ QED corrections that are included in $a_\mu^{\rm HVP, LO}$ by definition. 
About $90\%$ of the total HVP is comprised of the light-quark connected contribution, which therefore needs to be computed with subpercent precision. 

In the 2020 White Paper~\cite{Aoyama:2020ynm} the results of several collaborations published before the March 2020 deadline~\cite{\HVPlatticeref} were combined into a lattice HVP average with a total uncertainty of  approximately $2.6\%$, shown as the blue band in the right panel of Fig.~\ref{fig:compare}.  In 2021, a first lattice QCD result
with sub-percent uncertainty was published by the BMW collaboration \cite{Borsanyi:2020mff} (BMW20). Taken in isolation, the BMW20 result yields a reduced tension with the experimental average for $a_\mu$ of approximately
$1.5\sigma$, while being, at the same time, $2.1\sigma$ away from the reference result of Ref.~\cite{Aoyama:2020ynm} for the data-driven HVP calculation. However, the disagreement with the $R$-ratio approach becomes more pronounced
in the intermediate Euclidean time window $a_\mu^{\rm HVP,~LO,~W}$, introduced
in RBC/UKQCD18~\cite{Blum:2018mom}, where the integration over Euclidean time is restricted to an intermediate time region. For this quantity Ref.~\cite{Borsanyi:2020mff} finds a
result which is $3.7 \sigma$ above the corresponding data-driven evaluation.  Since the error of BMW20 is dominated by the uncertainty associated with the extrapolation to the continuum limit, it is of crucial importance to obtain results with similar precision from independent calculations employing different discretizations of the QCD action.

The Euclidean time windows allow for contributions from different Euclidean times to be studied separately. In lattice QCD calculations of the windows  one can disentangle statistical and systematic uncertainties, which affect the various Euclidean time regions differently.  Statistical noise and finite-volume uncertainties are most relevant at large Euclidean times, while discretization errors are typically enhanced at short distances.  The intermediate window, however, can be computed more easily at high precision in lattice QCD, which makes it a particularly attractive target for cross-checks between different lattice QCD calculations.  Indeed, already now multiple results with sub-percent precision are published for the isospin-symmetric light-quark connected contribution to this quantity~\cite{Blum:2018mom,Aubin:2019usy,Lehner:2020crt,Borsanyi:2020mff}, revealing that the BMW20 result is higher by $2.2\sigma$ compared to the one of RBC/UKQCD18~\cite{Blum:2018mom}, while consistent with the lattice results of Aubin19~\cite{Aubin:2019usy} and LM20~\cite{Lehner:2020crt}.
Furthermore, the Euclidean windows enable a powerful cross check by studying each window quantity (short-distance, intermediate, long-distance) separately in the continuum and infinite volume limits and comparing their sum to the direct evaluation of the total HVP contribution. 

At this point, additional sub-percent calculations of the intermediate Euclidean time window as well as the total HVP are crucial.  Several collaborations (including Aubin {\it et al.} \cite{Aubin:2021vej}, ETM \cite{Giusti:2021dvd}, FNAL/HPQCD/MILC \cite{Lahert:2021xxu,FermilabLattice:2021hzx}, Mainz \cite{Risch:2021hty}, and RBC/UKQCD) have on-going efforts for both quantities; however, due to the relative simplicity of $a_\mu^{\rm HVP,~LO,~W}$, we may expect additional results for it first.  The next generation of lattice QCD
results will also build on recent methodological advances made in the last years.  This includes the use of an exclusive state study for an improved long-distance computation (Mainz~\cite{Erben:2019nmx}, RBC/UKQCD~\cite{Bruno:2019nzm}, and FNAL/MILC~\cite{Lahert:2021xxu}).  Most groups plan to include gauge-field ensembles at smaller lattice spacings to test the continuum extrapolations, which is computationally very demanding and requires adequate computational resources. Methods are also being developed to control discretization effects specifically for the short-distance contribution~\cite{Harris:2021azd}. In addition, new theoretical insights, for example on the quark mass dependence of the light-quark contribution to $a_\mu^{\rm HVP}$ \cite{Colangelo:2021moe}, may improve control over certain systematic effects. Finally, as
lattice QCD calculations enter the sub-percent-precision era, several collaborations (FNAL/HPQCD/MILC and RBC/UKQCD) have started to implement blind analyses.

The Euclidean windows can also be evaluated straightforwardly in the data-driven approach.  Once tensions between the results from different lattice collaborations are resolved, detailed comparisons of results for the windows, as well as other sub quantities (see, for example, Ref.~\cite{Boito:2022rkw}), from lattice QCD and the data-driven approach will yield refined tests of the two approaches to HVP. In addition, assuming that any tensions between the two approaches are understood and resolved, windowed quantities may provide a useful strategy for combining lattice  and data-driven results to yield a better precision on the total $a_\mu^{\rm HVP,~LO}$ than each would by itself~\cite{Blum:2018mom}.

We expect that by the end of 2022, several new results for the intermediate window quantity and one or more additional sub-percent-precision results for the total HVP will be published.  The workshops organized by the Theory Initiative will continue to provide an open platform to facilitate further cross-checks, to define quality criteria for inclusion in the next iteration of the White Paper, and to develop a method average for lattice HVP based on detailed comparisons of individual contributions and subquantities to take into account any tensions between them and obtain a conservative estimate of the lattice HVP uncertainty.  The next Theory Initiative workshop~\cite{Higgs2022} will build on first steps towards this goal started at KEK in June 2021~\cite{KEK2021}.
Based on preliminary reports by several lattice QCD collaborations and assuming that any tensions between different lattice results are resolved, a lattice HVP average with $\le 0.5\%$ errors appears feasible by 2025.

\section{Data-driven and dispersive approach to HLbL}
\label{sec:dataeHLbL}

\begin{table}
\centering
\small 
\begin{tabular}{crrrr}
\toprule
  Contribution & Ref.~\cite{Prades:2009tw}  & 
  Refs.~\cite{Nyffeler:2009tw, Jegerlehner:2009ry} & 
  Ref.~\cite{Jegerlehner:2017gek} & 
  Ref.~\cite{Aoyama:2020ynm}\\
\midrule
 $\pi^0,\eta,\eta'$-poles & $ 114(13) $ & $ 99(16) $ & $ 95.45
 (12.40) $ & $ 93.8(4.0) $ 
\tabularnewline 

$\pi,K$-loops/boxes & $ -19(19) $ & $ -19(13) $ & $ -20(5) $ & 
 $ -16.4(2) $
\tabularnewline 
$S$-wave $\pi\pi$ rescattering & $ -7(7) $ & $ -7(2) $ & $-5.98(1.20) $ &  
 $ -8(1) $
\tabularnewline 
\midrule
subtotal & $88(24)$ & $73(21)$ & $69.5(13.4)$ & $69.4(4.1)$
\tabularnewline 
\midrule
scalars & $-$ & $-$ & $-$ &    
  \multirow{2}{*}{$\bigg\}\qquad -1(3)$} 
 \tabularnewline 

tensors & $-$ & $-$ & $ 1.1(1) $ & 
 \tabularnewline 

axial vectors & $ 15(10) $ & $ 22(5) $ & $ 7.55(2.71) $ &  
 $ 6(6) $ 
\tabularnewline 

~$u,d,s$-loops / short-distance~ & $-$ & $ 21(3) $ & $ 20(4) $
&   $ 15(10) $   
\tabularnewline \midrule
$c$-loop & $2.3$ & $-$ & $2.3(2)$
&   $ 3(1)$   
\tabularnewline 
\midrule
total & $ 105(26) $ & $ 116(39) $ & $ 100.4(28.2) $ &   
  $ 92(19) $ 
\tabularnewline 
\bottomrule
\end{tabular}
\caption{Comparison of the recommendation from Ref.~\cite{Aoyama:2020ynm} to two frequently used compilations for HLbL  from 2009 (``Glasgow
  consensus''~\cite{Prades:2009tw} and Jegerlehner/Nyffeler~\cite{Nyffeler:2009tw, Jegerlehner:2009ry}) and the recent update~\cite{Jegerlehner:2017gek}; in units of
  $10^{-11}$.} 
  \label{tab:compilations}
\end{table}

The organization of the phenomenological estimate of the HLbL contribution from Ref.~\cite{Aoyama:2020ynm} follows the same guiding principles as the data-driven evaluation of HVP: one considers the lowest-lying singularities of the HLbL tensor in the timelike region and explicitly estimates their contribution in a dispersive approach~\cite{Colangelo:2014dfa,Colangelo:2014pva,Pauk:2014rfa,Colangelo:2015ama}. That this approach is sensible is shown by the clear hierarchy among the contributions of different intermediate states based on their mass. This strategy needs to be supplemented by short-distance constraints (SDCs), which are relevant for large spacelike momenta and are imposed where applicable. While this procedure becomes significantly more complicated than the analog dispersion relation in the HVP case, it does allow for an, in principle, model-independent evaluation and thus provides a convenient framework for a data-driven approach. Moreover, results from lattice QCD can be used to further constrain the required input.   

The status according to Ref.~\cite{Aoyama:2020ynm} is summarized in Table~\ref{tab:compilations} in comparison to earlier compilations. The first panel shows the dominant contributions from pseudoscalar poles~\cite{Masjuan:2017tvw,Hoferichter:2018kwz,Gerardin:2019vio,Hoferichter:2018dmo}, boxes, and rescattering corrections~\cite{Colangelo:2017fiz,Eichmann:2019bqf,Colangelo:2017qdm}, yielding about $75\%$ of the total with well quantified uncertainties of $\approx 6\%$. In particular, for the $\pi^0$-pole contribution there is agreement among Canterbury approximants~\cite{Masjuan:2017tvw}, dispersion relations~\cite{Hoferichter:2018kwz,Hoferichter:2018dmo}, and lattice QCD~\cite{Gerardin:2019vio}. In this part of the evaluation, ongoing work thus mainly concerns a consolidation of the $\eta$, $\eta'$ poles, both using dispersion relations~\cite{Holz:2015tcg,Holz:2022hwz} and lattice QCD~\cite{Burri:2021cxr,Gerardin:2021fee} (see Refs.~\cite{Miramontes:2021exi,Stamen:2022uqh} for recent work on the box contributions).

The main uncertainty arises from the second panel in Table~\ref{tab:compilations}, which includes subleading contribution from higher intermediate states (approximated in terms of scalar, tensor, and axial-vector resonances~\cite{Pauk:2014rta,Danilkin:2016hnh,Jegerlehner:2017gek,Knecht:2018sci,Roig:2019reh}) and the implementation of SDCs~\cite{Melnikov:2003xd,Bijnens:2019ghy,Colangelo:2019uex,Colangelo:2019lpu}. A clear definition of individual narrow-resonance contributions and a distinction between these and the SDCs is at present affected by ambiguities~\cite{Colangelo:2017fiz,Danilkin:2021icn,Colangelo:2021nkr}, which yet need to be resolved or better understood. For this reason uncertainties in this panel were added linearly, as the errors in this category are potentially strongly correlated. Ongoing work is aimed at improving estimates of these subleading contributions, including SDCs at higher orders~\cite{Bijnens:2020xnl,Bijnens:2021jqo}, the implementation of these SDCs~\cite{Leutgeb:2019gbz,Cappiello:2019hwh,Masjuan:2020jsf,Ludtke:2020moa,Leutgeb:2021mpu,Colangelo:2021nkr}, and the evaluation of narrow resonances~\cite{Hoferichter:2020lap,Zanke:2021wiq,Danilkin:2021icn,Cappiello:2021vzi}.

Crucial ingredients in this program now concern the two-photon couplings of hadronic states in the $(1\text{--}2)\GeV$ region, most prominently of axial-vector resonances: unfortunately only limited experimental information is currently available for their transition form factors~\cite{Zanke:2021wiq,L3:2007obw,L3:2001cyf}; see also the discussion in Ref.~\cite{Aoyama:2020ynm}. New experimental results are expected in the future, e.g., from the two-photon program at BESIII~\cite{BESIII:2020nme}, and further constraints could be obtained from lattice QCD. 

In view of these ongoing developments, and the current error estimate of $\approx 20\%$ (based on a linear addition of uncertainties), it seems feasible to obtain a dispersive, data-driven evaluation of the HLbL contribution  
with $\le 10\%$ total uncertainty by 2025.

\section{Lattice QCD calculations of HLbL}
\label{sec:latticeHLbL}

In Ref.~\cite{Aoyama:2020ynm} the data-driven evaluation of HLbL scattering was combined with the first complete direct lattice QCD calculation performed by RBC/UKQCD~\cite{Blum:2016lnc,Blum:2019ugy} after cross-checks between the RBC/UKQCD and Mainz group for heavier pion mass were performed.  These cross-checks were facilitated by discussions during the Theory Initiative workshops in the preceding years~\cite{FNAL2017,KEK2018,UConn2018,Mainz2018,INT2019}.  The RBC/UKQCD calculation uses a finite-volume regulator for the photon (the QED$_{\rm L}$ prescription~\cite{Hayakawa:2008an}) and is based on gauge ensembles at the physical pion mass with chirally symmetric domain-wall fermions.  

An alternative infinite-volume photon method (QED$_\infty$) was proposed by the Mainz collaboration~\cite{Asmussen:2016lse} and refined by both the Mainz~\cite{Asmussen:2019act,Chao:2020kwq} and RBC/UKQCD~\cite{Blum:2017cer} collaborations, culminating in the publication of the second complete direct lattice calculation by the Mainz group~\cite{Chao:2021tvp} in 2021.  The Mainz calculation uses gauge ensembles with pion mass as low as 200 MeV with Wilson-clover fermions generated by the CLS effort.
At this point both lattice results are compatible with each other and with the data-driven result.

The two groups continue to improve their calculations.  RBC/UKQCD is focusing on a second result using QED$_\infty$ as well as improvements targeting a reduction of the statistical noise.  The Mainz group will continue towards adding data at physical pion mass.  In both cases, individual calculations at or below $10\%$ total uncertainty are feasible by 2025.  It is also expected that over the next years additional lattice collaborations may provide direct calculations of the  HLbL contribution as well.

\section{Conclusions and Outlook}
\label{sec:outlook}

As outlined in Secs.~\ref{sec:dataHVP}--\ref{sec:latticeHLbL}, it is reasonable to expect that by 2025 results for HVP and HLbL from two independent approaches will be available, each at or near the precision required to match the plans of the $g-2$ experiments. If for both HVP and HLbL data-driven and lattice determinations are found to be in good agreement, this will yield a SM prediction for $a_\mu$ with unprecedented precision, maximizing the discovery potential of the experimental efforts.

If, on the other hand, significant tensions between data-driven and lattice results are revealed, in particular for HVP, a continued effort will be needed in order to understand where these tensions arise and how sub quantities, such as Euclidean window quantities (see Sec.~\ref{sec:latticeHVP}), can help clarify the situation. It will also be important to explore in detail the connections between HVP, $e^+e^-$ cross sections and related low-energy parameters, as well as the hadronic corrections to the running of $\alpha$ and the global electroweak fit~\cite{Passera:2008jk,Crivellin:2020zul,Keshavarzi:2020bfy,Malaescu:2020zuc,Colangelo:2020lcg,Ce:2022eix}. Further insights into these connections will be provided by another complementary method for HVP, which is expected to become available over the next years at the MUonE experiment~\cite{CarloniCalame:2015obs,Abbiendi:2016xup,MUonE:LoI,Banerjee:2020tdt}, via a space-like measurement of HVP in muon--electron scattering, with recent work addressing both the experimental realization~\cite{Ballerini:2019zkk,Abbiendi:2019qtw,Abbiendi:2021xsh,Abbiendi:2022oks} and theory corrections~\cite{Mastrolia:2017pfy,DiVita:2018nnh,Fael:2018dmz,Alacevich:2018vez,Fael:2019nsf,CarloniCalame:2020yoz,Banerjee:2020rww,Bonciani:2021okt,Budassi:2021twh,Nesterenko:2021byp,Balzani:2021del,Fael:2022rgm,Greynat:2022geu,Budassi:2022kqs}.
Hadronic $\tau$-decay data can, in principle, also be used to evaluate HVP, which, however, requires a determination of the needed isospin correction~\cite{Alemany:1997tn}. While phenomenological estimates of this correction are not sufficiently quantified, it may be possible to compute it reliably in lattice QCD~\cite{Bruno:2018ono}. If precise lattice results for the isospin correction became available, then $\tau$-decay data could be used to provide interesting cross checks for HVP. We note that precise measurements of $\tau$-decay spectral functions are expected from Belle II in the coming years.  

To make optimal use of all these developments, the Muon $g-2$ Theory Initiative continues its work, with two workshops~\cite{HVP2020,KEK2021} held after the completion of Ref.~\cite{Aoyama:2020ynm} and the next plenary meeting to be held in September 2022~\cite{Higgs2022}. At this meeting, concrete plans for White Paper updates will be discussed, with a timeline depending on the availability of new results especially regarding lattice-QCD calculations of HVP. 
A main update is anticipated for 2023 and will include any new available results as well as a method average for lattice HVP and HLbL. Most crucially, the Muon $g-2$ Theory Initiative will continue to facilitate interactions among the different groups and communities involved in the SM prediction of the anomalous magnetic moment of the muon, including experiment, phenomenology, and lattice QCD.   

Beyond 2025, the experimental $g-2$ program will continue at  J-PARC~\cite{Abe:2019thb}, with a completely independent experimental technique, 
and potentially even further, with the  High-Intensity Muon Beams (HIMB) project at PSI~\cite{Aiba:2021bxe} and the Muon Campus at Fermilab~\cite{FutMuon2021}. To fully exploit the final E989 Fermilab result and keep up with these future experiments, it is evident that a sustained effort of further improving the SM prediction for $a_\mu$ is needed beyond 2025, even if all near-term goals laid out above can be achieved. For the data-driven evaluation of HVP, this requires a sustained program at $e^+e^-$ machines (at Belle II, BES III, and Novosibirsk), together with the calculation of higher-order radiative corrections and the development of MC generators at NNLO accuracy. For HLbL scattering, both improved input data and further theory improvements will be needed.
For the lattice HVP and HLbL calculations, systematic and statistical uncertainties can be improved significantly with large-scale access to future leadership-class computing facilities. Further methodological developments, for example, to more efficiently generate gauge field ensembles  with small lattice spacings and large volumes (see, for example, Refs.~\cite{Albergo:2021bna,Hackett:2021idh,Foreman:2021ljl,Nguyen:2021zgx,Boyda:2022nmh}) as well as improved statistical noise reduction (see, for example, Refs.~\cite{DallaBrida:2020cik,Detmold:2021ulb,Giusti:2021qhk}) would further enhance the impact of these future computational resources on the precision of the lattice QCD results. 
The Muon $g-2$ Theory Initiative will continue to coordinate efforts along these lines, to ensure maximal return on the investments made in the experimental $g-2$ program.   

\section*{Acknowledgments}

\begin{sloppypar}
We thank F.~Ignatov, A.~Nesterenko, and A.E.~Radzhabov for comments and support. 
The work in this contribution was supported 
by Agence Nationale de la Recherche	(ANR-19-CE31-001),
by CERCA program of the Generalitat de Catalunya (2017 SGR 1069), 
by Conacyt 'Paradigmas y Controversias de la Ciencia 2022' (project number 319395), 
by Danmarks Frie Forskningsfond under Grant Number 8021-00122B, 
by Deutsche Forschungsgemeinschaft Collaborative Research Centers CRC~1044, CRC~110, and under Grant Numbers DE 839/2-1, HA 9289/1-1, HI 2048/1-2 (Project No.\ 399400745), Prisma Cluster for Excellence PRISMA$^+$ EXC2118/1,
by DST, Govt.~of India	INSPIRE Faculty Fellowship (grant number IFA16-PH170),
by European Research Council under the European Union's Horizon 2020 research and innovation programme under Grant Agreement Number 771971-SIMDAMA, under the Marie Sk\l{}odowska-Curie grant agreement numbers 860881-HIDDEN, 894103, under grant number	754510 (EU, H2020-MSCA-COFUND2016), by European Union STRONG 2020 project under Grant Agreement Number 824093, 
by the Excellence Initiative of Aix-Marseille University - A*MIDEX, a French ``Investissements d'Avenir'' program, through grants AMX-18-ACE-005 and AMX-19-IET-008-IPhU, and by the French National Research Agency under contract ANR-20-CE31-0016, 
by FWF (grant numbers I 3845-N27 and W 1252-N27), 
by Fundaci\'on Marcos Moshinsky	(C\'atedra Marcos Moshinsky 2020),
 by the Japan Society for the Promotion of Science under Grant Numbers KAKENHI-16K05317, 20K03926, 20H05625,   
by the National Research Foundation of South Africa, 
by the Natural Sciences and Engineering Research Council of Canada, 
by Spanish Ministry of Science, Innovation and Universities 	(PID2020-112965GB-I00/AEI/10.13039/501100011033), 
by the Swedish Research Council under Grant Numbers 2016-05996, 2019-03779, 
by the Swiss National Science Foundation under Grant Numbers  PCEFP2\_181117, PCEFP2\_194272, PZ00P2\_193383, 200021\_200866, 200020\_175791, 
by the UK Science and Technology Facilities Council (STFC) under Grant Numbers ST/P000630/1, ST/T000988/1, ST/S000925/1,
by the U.S.\ Department of Energy, Office of Science, Office of High Energy Physics under Award Numbers DE-SC0007983, DE-SC0010005, DE-SC0010339, DE-SC0011941, DE-SC0012704, DE-SC0013682, DE-SC0015655, by the U.S.\ Department of Energy, Office of Science, Office of Nuclear Physics under Award Numbers DE-AC02-07CH11359, and
by the U.S.\ National Science Foundation under Grant Numbers PHY-1748958
This document was prepared by the Muon $g-2$ Theory Initiative using the resources of the Fermi National Accelerator Laboratory (Fermilab), a U.S.\ Department of Energy, Office of Science, HEP User Facility.  Fermilab is managed by Fermi Research Alliance, LLC (FRA), acting under Contract No.\ DE-AC02-07CH11359.
\end{sloppypar}

\bibliographystyle{apsrev4-1_mod}
\bibliography{main.bib}

\end{document}